\documentclass[aps,prb,twocolumn,superscriptaddress,showpacs,floatfix]{revtex4}

\usepackage{epsfig}

\begin{document}

\title{Anisotropic superconducting strip in an oblique magnetic
field}

\author{E.~H.~Brandt}
\affiliation{Max-Planck-Institut f\"ur Metallforschung,
   D-70506 Stuttgart, Germany}

\author{G.~P.~Mikitik}
\affiliation{Max-Planck-Institut f\"ur Metallforschung,
   D-70506 Stuttgart, Germany}
\affiliation{B.~Verkin Institute for Low Temperature Physics
   \& Engineering, Ukrainian Academy of Sciences,
   Kharkov 61103, Ukraine}

\date{\today}

\begin{abstract}
The critical state of a thin superconducting strip in an oblique
applied magnetic field $H_a$ is analyzed without any restrictions
on the dependence of the critical current density $j_c$ on the
local magnetic induction ${\bf B}$. In such a strip, $j_c$ is not
constant across the thickness of the sample and differs from
$J_c/d$, where $J_c$ is the critical sheet current. It is shown
that in contrast to the case of ${\bf B}$-independent $j_c$, the
profiles $H_z(x)$ of the magnetic-field component perpendicular to
the strip plane generally depend on the in-plane component
$H_{ax}$ of the applied magnetic field $H_a$, and on how $H_a$ is
switched on. On the basis of this analysis, we explain how and
under what conditions one can extract $j_c({\bf B})$ from the
magnetic-field profiles $H_z(x)$ measured by magneto-optical
imaging or by Hall-sensor arrays at the upper surface of the
strip.
\end{abstract}


\maketitle

\section{Introduction}

One of the methods for analyzing flux-line pinning in high-$T_c$
superconductor thin platelets is the measuring of magnetic-field
profiles at the upper surface of these superconductors when they
are placed in a perpendicular external magnetic field; see, e.g.,
the recent review \cite{Kr} and references therein. These profiles
enable one to reconstruct the spatial distribution of the sheet
currents $J$ flowing in the sample \cite{Kr,R,JH,Gr,W,J,G,Joo}
(the sheet current $J$ is the current density integrated over the
thickness of the sample), and hence to extract some information on
the critical currents in the superconductor.\cite{ita} The
magnetic-field profiles are measured either with magneto-optical
imaging, \cite{In,Kob}$^,$\cite{Kr} or with Hall-sensor
arrays.\cite{Zel} It is clear that the investigation of the
profiles not only in a perpendicular but also in an oblique
magnetic field could, in principle, provide additional information
on vortex pinning in thin flat superconductors.

In this paper, to explain the main ideas, we shall consider a
sample of the simplest shape: Let a thin superconducting strip
fill the space $|x| \le w$, $|y| < \infty$, $|z| \le d/2$ with
$d\ll w$. A constant and homogeneous external magnetic field $H_a$
is applied in the $x$-$z$ plane at an angle $\theta_0$ to the $z$
axis ($H_{ax}=H_a\sin\theta_0$, $H_{ay}=0$,
$H_{az}=H_a\cos\theta_0$). We also assume that there is no surface
pinning, the thickness of the strip, $d$, exceeds the London
penetration depth, and the lower critical field $H_{c1}$ is
sufficiently small so that we may put for the magnetic induction
$B= \mu_0 H$.

In our recent paper, \cite{mbi} a superconducting strip in an
oblique magnetic field was analyzed in the framework of the Bean
model (when the bulk critical current density $j_c$ is {\it
independent} of the local magnetic induction ${\bf B}$), and
several interesting properties of the critical state for such
``nonsymmetric'' directions of $H_a$ were established. In
particular, it was shown that the final critical state depends on
how the applied magnetic field is switched on, e.g., at a constant
tilt angle $\theta_0$, or first its $z$ and then its $x$
component, or vise versa. However, differences between these three
critical states occur only in the central region of the strip
where a flux-free core exists, i.e., where $H_z(x)$ at the surface
of the strip is practically zero. Moreover, to leading order in
the small parameter $d/w$, for all the scenarios the sheet current
$J(x)$ is the same in this region of the sample, and it coincides
with the sheet current at $H_{ax}=0$. In the region of the strip
where $H_z\neq 0$, the sheet current is equal to its critical
value $J_c=j_cd$ and is also independent of $H_{ax}$ and of how
$H_a$ is switched on. Thus, when the Bean model holds, the
magnetic-field profiles depend only on $H_{az}$, the differences
between the states do not manifest themselves in the profiles,
\cite{c1} and hence measurements of the $H_z$ profiles in inclined
magnetic fields do not provide new information on flux-line
pinning. In this case, to find $j_c$, it is sufficient to extract
the critical sheet current $J_c$ from the $H_z$ profiles at
$H_{ax}=0$ in the region with a nonzero $H_z$ and to use the
relationship $j_c=J_c/d$.

However, when the critical current density $j_c$ depends on ${\bf
B}$, e.g., on $|B|$ or on the angle between the local direction of
${\bf B}$ and the $z$ axis or on both, the situation is different.
In this case $j_c$ is not constant across the thickness of the
strip since ${\bf B}$ and thus $j_c$ change with $z$ in a strip of
finite thickness, and the critical value of the sheet current
$J_c$ is not simply $j_cd$. In such cases the problem arises how
to determine $j_c({\bf B})$ from the magnetic-field profiles. In
the present paper we address just this problem. As it was shown in
Ref.~\onlinecite{ani} (see also below), at {\it any} dependence of
$j_c$ on ${\bf B}$ the critical state problem for a thin strip of
finite thickness can be reduced to the problem of an infinitely
thin strip with some effective dependence of the critical sheet
current $J_c$ on the $z$ component of the local magnetic
induction, $\mu_0H_z$. Thus, the problem can be formulated as the
reconstruction of $j_c({\bf B})$ from the measured $J_c(H_z)$.
This $J_c(H_z)$ depends also on the applied $H_{ax}$, and hence
some additional information on flux line pinning can, in
principle, be extracted from the $H_z$ profiles measured in an
oblique magnetic field.

In this paper, to provide a basis of treating the reconstruction
problem, we first consider the critical-state problem for the
superconducting strip in an oblique magnetic field, without
assuming any restrictions on the dependence $j_c({\bf B})$. In
contrast to the case of ${\bf B}$-independent $j_c$, critical
states in such a strip depend on how the external magnetic field
is switched on, not only in the region of the strip where the
flux-free core occurs but also in the region penetrated by $H_z$.
In particular, the profiles $H_z(x)$ at the upper surface of the
strip generally depend on $H_{ax}$ and on how $H_a$ is switched
on. As an example, in this paper we in detail analyze the case
when $j_c$ depends only on the angle $\theta$ between the local
direction of ${\bf H}$ and the $z$ axis, $j_c=j_c(\theta)$.
Finally, we discuss the problem of the reconstruction of $j_c({\bf
B})$ from experimental magnetic-field profiles.

\section{Critical state in a thin strip}

Let us consider the critical state in a thin strip with an
arbitrary dependence $j_c({\bf B})$. In the specific case
$H_{ax}=0$, this problem was solved in Ref.~\onlinecite{ani}. Here
we generalize the approach of that paper to the case of an oblique
magnetic field. We assume that the field $H_a$ increases
monotonically and imply that one of the following three scenarios
occurs: In the first scenario the magnetic field increases under
the condition $(H_{ax}/H_{az})=\tan\theta_0=\ $const. In the
second scenario the field $H_{az}$ is switched on first and then
one switches on $H_{ax}$, while for the third scenario the
sequence of switching on $H_{ax}$ and $H_{az}$ is opposite. The
final magnetic field in all these cases is the same,
$H_{ax}=H_a\sin\theta_0$, $H_{ay}=0$, $H_{az}=H_a\cos\theta_0$.

To the leading order in $d/w$, $H_z$ is independent of $z$ inside
the strip, and the Biot-Savart law gives
\begin{equation} \label{1}
 H_z(x)=H_{az}+{1 \over 2\pi}\!\int_{-w}^w\!\!
 { J(v)\, dv \over v-x}  \,,
\end{equation}
where the sheet current $J$ is the current density $j_y$
integrated over the thickness of the strip,
\[
 J(x) \equiv \int_{-d/2}^{d/2} j_y(x,z)dz .
\]
When $H_a$ increases, the penetration of the $z$ component of the
magnetic field into the superconducting strip is found from the
following equations: One has
\begin{equation}\label{2}
 J(x)=-{x\over |x|}J_c[H_z(x)]
\end{equation}
in the regions of the strip where $H_z$ changes as compared to the
previous state that existed at $H_a-\delta H_a$, while in the
regions where the sheet current is less than its critical value
$J_c[H_z(x)]$, $H_z$ remains unchanged,
\begin{equation}\label{3}
 \delta H_z(x)=0.
\end{equation}
In the simplest case, Eq.~(\ref{2}) holds in the two regions $a\le
|x| \le w$ where $H_z$ differs from zero, while in the central
region $|x|\le a$, one has
\begin{equation}\label{3a}
 H_z=0.
\end{equation}
These equations, in fact, describe the critical state problem for
an infinitely thin strip with some effective dependence of the
critical sheet current $J_c$ on $H_z$. The point $x=a$ where $H_z$
vanishes [or in the general case the boundaries separating regions
of validity of Eqs.~(\ref{2}) and (\ref{3})] is found by solving
Eqs.~(\ref{1})-(\ref{3a}).

The function $J_c(H_z)$ can be found as follows: \cite{ani} Taking
into account that in a thin strip the critical state gradient of
the magnetic field develops predominantly across the thickness of
the sample, we can write the equation ${\rm rot}{\bf H}={\bf j}$
in the form
\begin{equation} \label{4}
 {\partial H_x(x,z)\over \partial z}= j_y(x,z),
\end{equation}
neglecting the derivative $\partial H_z(x,z)/ \partial x$. This
approximation is true  to the leading order in $d/w$. Here
$j_y(x,z)=\pm j_c(H_x,H_z)$ is the current density in the critical
state. Solving Eq.~(\ref{4}), one finds a relation that determines
the function $J_c(H_z)$ implicitly,
\begin{equation} \label{5}
 d= \!\! \int_{H_x^-}^{H_x^+}\!\!{dh
 \over j_c(h,H_z)} ,
\end{equation}
where $H_x^-=H_{ax}-0.5J_c(H_z)$, and $H_x^+=H_{ax}+0.5J_c(H_z)$
are the $x$ component of the magnetic field at the lower and the
upper surfaces of the strip, $H_x(x,\pm d/2)$. The function
$J_c(H_z)$ found from Eq.~(\ref{5}) generally depends on the
parameter $H_{ax}$, and only within the simple Bean model when
$j_c$ is independent of ${\bf H}$, equation (\ref{5}) yields
$J_c=j_cd$ for any $H_{ax}$. At $H_{ax}=0$, formula (\ref{5})
coincides with formula (12) of Ref.~\onlinecite{ani}.

In the regions of the strip where $|J|<J_c$, either the flux-free
core occurs, or there is a boundary which is almost parallel to
the strip plane and at which the critical current density
$j_y(x,z)$ changes its sign; see below. Knowing $J(x)$ and using
Eq.~(\ref{4}), one can calculate the shape of the flux-free core
and this boundary, i.e., one can find the critical state across
the thickness of the strip. The sign of $j_y$ above and below the
boundary, as well as the sign on the right hand side of
Eq.~(\ref{2}), follows from the prehistory of the critical state.
\cite{crst}

\section{$j_c$ depends only on the direction of {\bf H}}

Anisotropic pinning is typical of anisotropic superconducting
materials, and it occurs in many superconductors. Only for the
case of weak collective pinning by point defects is the critical
current density $j_c$ independent of the magnetic induction in the
so-called single vortex pinning regime or does it depend on $B_z$
in the small bundle pinning regime. \cite{bl} In this case
$J_c=j_cd$, and $j_c$ can be directly found from the critical
value of the measured sheet current $J_c$. However, anisotropy of
pinning in an anisotropic material is likely if the defects cannot
be considered as point pinning centers or if the pinning is not
weak. Anisotropic pinning also occurs in layered superconductors.
\cite{bl} Besides this, it is clear that any extended defects,
e.g., twin boundaries or columnar defects, lead to an anisotropy
of pinning even in isotropic materials.

Let us consider the case when $j_c$ depends only on $\theta$,
$j_c=j_c(\theta)$, where $\theta$ is the angle between the local
direction of ${\bf H}$ and the $z$ axis. In other words, we
neglect a possible dependence of $j_c$ on $|H|$ here. When the
characteristic scale $H_0$ of the $|H|$ dependence considerably
exceeds
the characteristic field in the critical state of the strip,
$j_cd$, this approximation is well justified since one has
$j_c(\theta,|H|)\approx j_c(\theta,0)$ in such a strip.
Hence, this approximation is quite reasonable for not too thick
samples with $d\ll H_0/j_c$. On the other hand, if $j_c$ depends
on $\theta$ and $|H|$ separately rather than on the combination
$|H|\cos\theta$, one cannot neglect the angular dependence of
$j_c$ even in thin samples since the characteristic angles
$\theta$ in the strip are of the order of $j_cd/H_z$ and
essentially change at $H_z\sim j_cd$.

In the case $j_c=j_c(\theta)$ with the symmetry
$j_c(-\theta)=j_c(\theta)$, one can reconstruct the dependence of
the critical current density $j_c$ on $\theta$ from the function
$J_c(H_z)$ obtained at $H_{ax}=0$. Indeed, differentiating
Eq.~(\ref{5}) with respect to $H_z$, we have at $H_{ax}=0$,
\cite{ani}
     \begin{eqnarray} \label{7}
 j_c(\theta) d = J_c(H_z) -H_z{d J_c(H_z) \over d H_z} \,, \nonumber \\
  \tan \theta   = { J_c(H_z) \over 2 H_z } \,.
     \end{eqnarray}
These formulas give the dependence $j_c(\theta)$ in parametric
form, with $H_z$ being the parameter. Inserting this parametric
form of $j_c(\theta)$ into Eq.~(\ref{5}), we arrive at equations
determining $J_c(H_z)$ at $H_{ax}\neq 0$,
     \begin{eqnarray}  \label{8}
 {0.5J_c(H_z,H_{ax})+H_{ax}\over H_z}&=&{J_c(t_+,0)\over 2t_+},
 \nonumber \\
 {|0.5J_c(H_z,H_{ax})-H_{ax}|\over H_z}&=&{J_c(t_-,0)\over 2t_-},
 \nonumber \\
 {1\over H_z}&=&{1\over 2t_+}+\sigma {1\over 2t_-},
     \end{eqnarray}
where $\sigma$ is the sign of $[0.5J_c(H_z,H_{ax})-H_{ax}]$,
$J_c(H_z,H_{ax})$ denotes $J_c(H_z)$ at a given value of $H_{ax}$,
and hence $J_c(t,0)$ is the sheet current at $H_{ax}=0$. Note that
at $H_{ax}=0$ one has $t_+=t_-=H_z$, and Eqs.~(\ref{8}) reduce to
$J_c=J_c(H_z,0)$ as it should be. From the three Eqs.~(\ref{8})
for the three unknown variables $J_c(H_z,H_{ax})$, $t_+$, $t_-$
($t_+$ and $t_-$ are auxiliary variables that do not appear in the
final results) one finds $J_c(H_z)$ at $H_{ax}\neq 0$. \cite{c2}
The magnetic field profiles in oblique applied field are then
obtained by inserting this effective law $J_c(H_z,H_{ax})$ into
Eqs.~(\ref{1})- (\ref{3a}), and by solving these integral
equations by either a static iterative method, or more
conveniently by a dynamic method. \cite{EH}

  \begin{figure}  
\includegraphics[scale=.45]{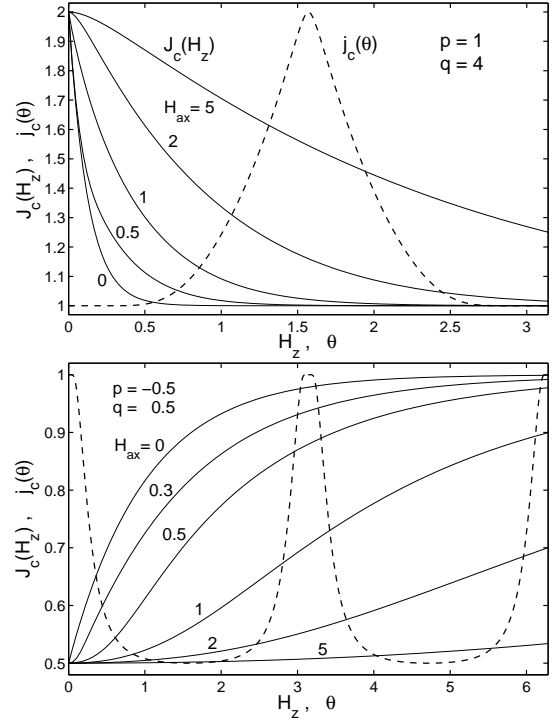}
\caption{\label{fig1} Angular dependence of the critical current
density $j_c(\theta)$, Eq.~(\ref{10}), (dashed lines) for $p=1$,
$q=4$ (top, the maximum of $j_c$ is at $\theta=\pi/2$) and for
$p=-0.5$, $q=0.5$ (bottom, the maximum of $j_c$ is at $\theta=0$).
The solid lines show the corresponding sheet currents $J_c(H_z)$
at $H_{ax}=0$, Eq.~(\ref{9}), and at $H_{ax}=0.3$, $0.5$, $1$,
$2$, $5$ from Eqs.~(\ref{8}), (\ref{9}). The current density is
measured in units of $j_c(0)$, while $J_c$ and $H_z$ are in units
of $j_c(0)d$.
 } \end{figure}  

We now consider typical examples of the dependence $j_c(\theta)$.
Let at $H_{ax}=0$ the following dependence $J_c(H_z)$ be extracted
from some experimental magneto-optics data:
    \begin{eqnarray}  \label{9}
    J_c(H_z)=J_c(H_z,0)= j_c(0) d \left [ 1 + p\, \exp\left
    (-{q \,H_z \over H_{cr}}\right ) \right ] \,,
    \end{eqnarray}
where $H_{cr} = j_c(0) d /2$, while $j_c(0)$ and the dimensionless
$p$ and $q$ are constants [$j_c(0)$ and $q>0$]. Using
Eqs.~(\ref{7}), one can easily verify that the corresponding
angular dependence of the critical current density takes the form:
     \begin{eqnarray}  \label{10}
     j_c(\theta) =j_c(0)\left [ 1 + p \, (1+ q\, t)
     \exp(- q\, t)\right ] \,, \nonumber \\
     \tan \theta = t^{-1}\left [ 1 + p\,
     \exp(- q\, t)\right ] \,,
     \end{eqnarray}
where $t$ is a curve parameter with range $0 \le t \le \infty$
equivalent to $\pi/2 \ge \theta \ge 0$. This dependence
$j_c(\theta)$ is presented in Fig.~\ref{fig1} together with the
function $J(H_z)$, Eq.~(\ref{9}). Note that the character of this
dependence $j_c(\theta)$ is typical of layered high-$T_c$
superconductors when $p>0$ and of superconductors with columnar
defects when $p<0$. Using Eqs.~(\ref{8}) and (\ref{9}), we find
the dependences of $J_c$ on $H_z$ for $H_{ax}\neq 0$,
Fig.~\ref{fig1}. We consider now $H_z$ profiles for various types
of pinning and for the different scenarios of switching on $H_a$.

 \begin{figure}  
\includegraphics[scale=.50]{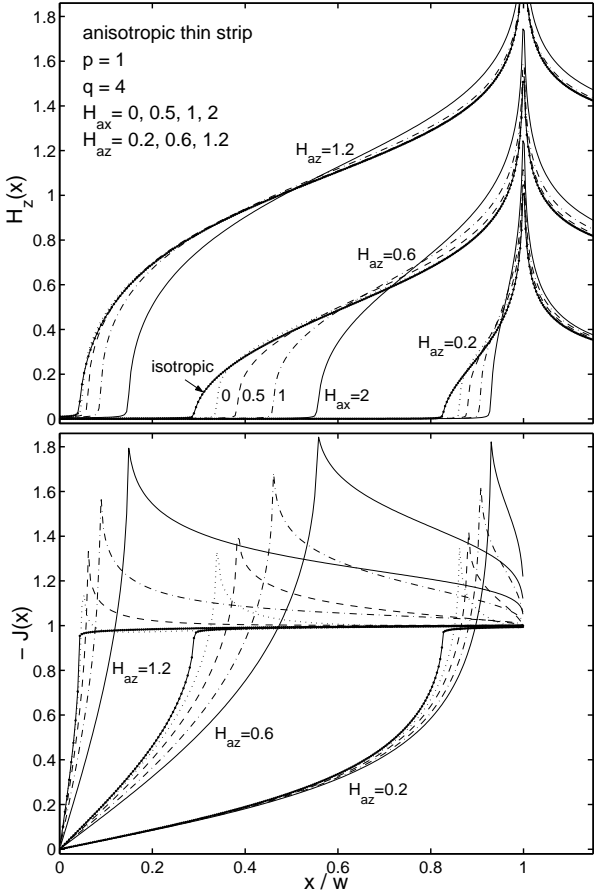}
\caption{\label{fig2} Spatial profiles of the perpendicular field
component  $H_z(x)$ (upper plot) and of the sheet current $J(x)$
(lower plot) of a thin strip with anisotropic pinning described by
model (\ref{10}) with $p=1$, $q=4$, see  Fig.~1. $H_{ax}$ is
switched on first and then $H_{az}$ (third scenario). The various
curves correspond to increasing applied field $H_{az}=0.2$, $0.6$,
$1.2$. The magnetic fields and the sheet current $J$ are in units
of $j_c(0)d$. The dotted, dashed, dot-dashed, and solid curves are
for $H_{ax}=0$, 0.5, 1, and 2, respectively. For comparison, the
solid curves with dots (indicating the grid) show the profiles for
isotropic pinning ($p=0$). The dotted lines also correspond to the
opposite sequence of switching on $H_{az}$ and $H_{ax}$ (second
scenario).
 } \end{figure}  

\subsection{Layered superconductors, $p>0$}

\begin{figure}  
\includegraphics[scale=.50]{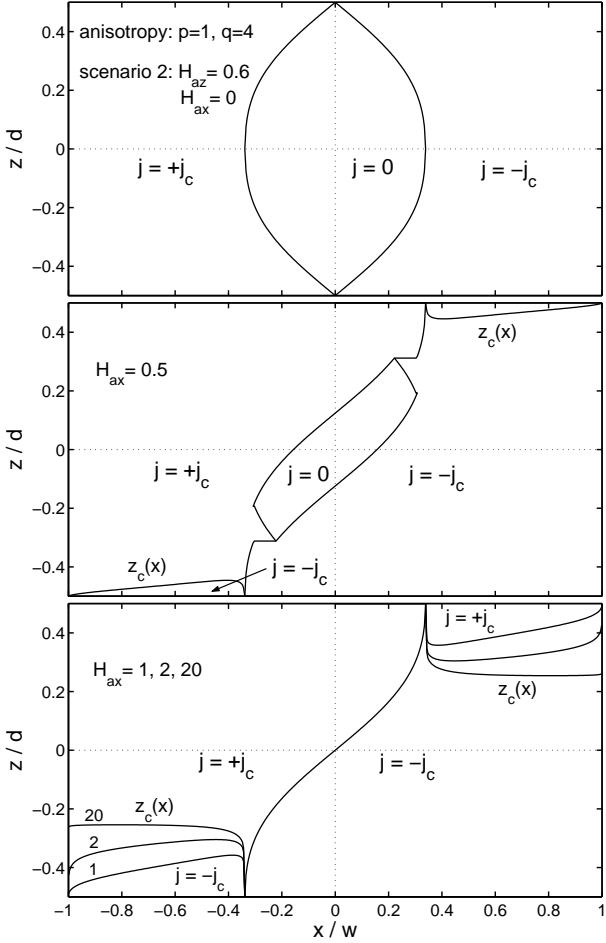}
\caption{\label{fig3} Current fronts and flux-free core in a thin
strip with anisotropic pinning described by model (\ref{10}) with
$p=1$, $q=4$. The perpendicular magnetic field $H_{az}=0.6$ is
applied first and then the in-plane component $H_{ax}$ is
increased from zero to $H_{ax}=0.5$, $1$, $2$, $20$ (second
scenario). At $|x|\le a= 0.339w$, the flux-free core and the lines
separating the regions of the strip with opposite directions of
currents are plotted, using \cite{c4} the results of
Ref.~\onlinecite{mbi}. The new boundary $z=z_c(x)$ which occurs at
$a \le |x| \le w$ is calculated from Eqs.~(\ref{12}). This
boundary remains practically unchanged when $H_{ax}\ge 8$. The
magnetic fields are measured in units of $j_c(0)d$.
 } \end{figure}

For layered superconductors 
($p>0$), we begin our analysis with the third scenario when
$H_{ax}$ is switched on first and then $H_{az}$ is applied.
Appropriate magnetic-field profiles obtained with Eqs.~(\ref{1}),
(\ref{2}), (\ref{3a}) and $J_c(H_z,H_{ax})$ of Fig.~1 are shown in
Fig.~\ref{fig2}. Note that for the third scenario one has
$|J|=J_c$ at any point $x$ where $H_z(x)\neq 0$, and the
penetration depth of $H_z$ into the strip, $w-a$, depends on
$H_{ax}$. Another situation occurs when $H_{ax}$ and $H_{az}$ are
switched on in the opposite sequence (the second scenario). After
switching on $H_{az}$, the magnetic-field profiles are given by
the dotted lines in Fig.~2. The subsequent increase of $H_{ax}$
{\it does not change} either $J(x)$ or $H_z(x)$. \cite{c} Indeed,
as seen from Fig.~1, $J_c(H_z)$ increases with increasing
$H_{ax}$. Hence, if $J(x)$ were equal to its critical value
$J_c[H_z(x)]$, the penetration depth of the applied field $H_{az}$
would decrease as compared to $w-a$ for the dotted lines of
Fig.~2, and flux lines near the points $x=\pm a$ would have to
move against the Lorenz forces directed towards the center of the
strip, which is impossible. Thus, for the second scenario the
sheet current is {\it less} than its critical value for all points
$x$. This means that although the fully penetrated critical state
develops across the thickness of the strip in the regions
penetrated by $H_z$, there is a boundary $z=z_c(x)$ at which the
critical current density $j_y(x,z)$ changes its sign,
Fig.~\ref{fig3}, and so the magnitude of sheet current $|J|$ is
less than $J_c$. It follows from Eq.~(\ref{4}) that the boundary
$z_c(x)$ is determined by the formula,
\begin{equation} \label{11}
 2z_c(x)= -\!\! \int_{H_x^-}^{H_x^+}\!\!{dh
 \over j_c(h,H_z)} ,
\end{equation}
where $H_x^-=H_{ax}-0.5J(x)$, and $H_x^+=H_{ax}+0.5J(x)$ are the
$x$ component of the magnetic field at the lower and the upper
surfaces of the strip at the point $x$, while the sheet current
$J(x)$ coincides with $J_c[H_z(x)]$ calculated at $H_{ax}=0$,
i.e., $J(x)=-(x/|x|)J_c[H_z(x),0]$. Inserting the parametric form
of $j_c(\theta)$, Eq.~(\ref{7}), into Eq.~(\ref{11}), we arrive at
the following equations for $z_c$:
     \begin{eqnarray}  \label{12}
 {0.5J_c(H_z,0)+H_{ax}\over H_z}={J_c(u_+,0)\over 2u_+},
 \nonumber \\
 {|0.5J_c(H_z,0)-H_{ax}|\over H_z}={J_c(u_-,0)\over 2u_-},
 \nonumber \\
 {2z_c(x)\over d}={{\rm sgn}(x) H_z \over 2}
 \left ({1\over u_+}+\sigma {1\over u_-}\right ).
     \end{eqnarray}
Here $u_+$ and $u_-$ are auxiliary variables; $\sigma$ is the sign
of $[0.5J_c(H_z,0)-H_{ax}]$; $H_z=H_z(x)$ is the magnetic-field
profile at $H_{ax}=0$, and ${\rm sgn}(x)=\pm 1$ for $x>0$ and
$x<0$, respectively.

For the first scenario of switching on $H_a$ when $H_{az}$ and
$H_{ax}$ increase simultaneously, the magnetic-field profiles
calculated for the same final values of $H_{az}$ and $H_{ax}$ as
in Fig.~\ref{fig2} coincide with the profiles for the third
scenario shown in this figure.

\subsection{Superconductors with columnar defects, $p<0$}

\begin{figure}  
\includegraphics[scale=.50]{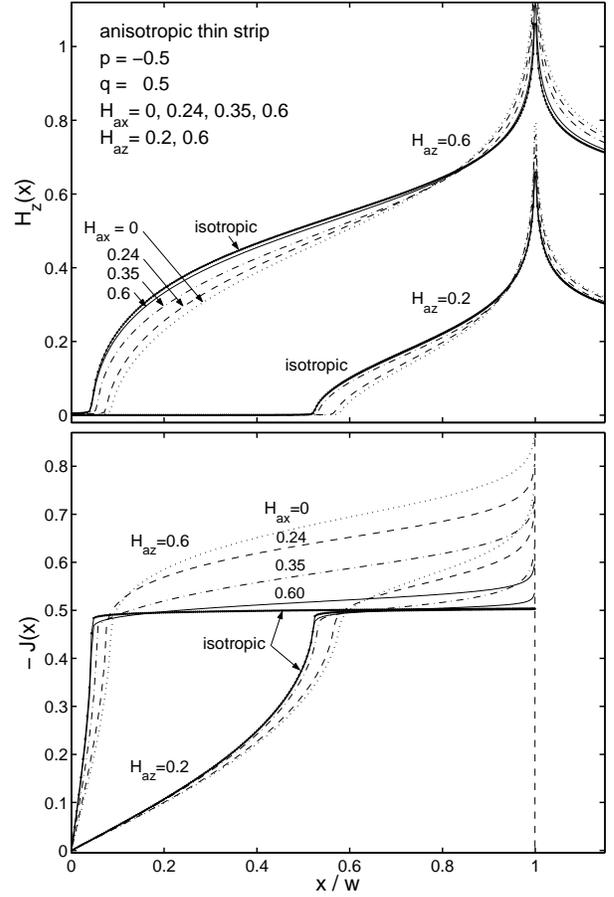}
\caption{\label{fig4} Spatial profiles of the perpendicular field
component  $H_z(x)$ (upper plot) and of the sheet current $J(x)$
(lower plot) of a thin strip with anisotropic pinning described by
model (\ref{10}) with $p=-0.5$, $q=0.5$, see  Fig.~1. The various
curves correspond to increasing applied field $H_{az}=0.2$, $0.6$.
The magnetic fields and the sheet current $J$ are in units of
$j_c(0)d$. The dotted, dashed, dot-dashed, and solid curves are
for $H_{ax}=0$, $0.24$, $0.35$, and $0.6$, respectively. At any
$H_{ax}>0.6$ the profiles practically coincide with those at
$H_{ax}=0.6$, while at $H_{ax}<0.2$ the profiles are very close to
that at $H_{ax}=0$. The profiles at $H_{ax}=0.6$ also almost
coincide with the profiles for isotropic pinning with
$j_c(\theta)=j_c(\pi/2)$ (solid curves with dots indicating the
used grid). The profiles $H_z(x)$ and $J(x)$ coincide for the
three scenarios of switching on $H_a$.
 } \end{figure}

Consider now a superconductor with columnar defects directed along
the $z$ axis. To model the sheet current $J(H_z)$ in such
superconductors, one can use Eq.~(\ref{9}) with $p<0$, Fig.~1. In
Fig.~\ref{fig4} we present the appropriate magnetic-field
profiles.  Note that these profiles coincide for the three
scenarios of switching on $H_a$ (neglecting the effect of flux
creep which can be different for different scenarios). However,
the profiles and the penetration depth of $H_z$ depend on
$H_{ax}$. When $H_{ax}$ increases, the profiles tend to those of
superconductors without columnar defects [it is evident from
Fig.~1 that pinning without these defects is implied to be
isotropic and to be described by the Bean model with constant
$j_c(\theta)=j_c(\pi/2)$]. The change of the profiles occurs in a
rather narrow interval of $H_{ax}$. This result can be understood
from the following simple considerations: Although the columnar
defects increase pinning in the strip, this increase occurs mainly
when the angle $\theta$ of a flux line does not exceed the
so-called trapping angle $\theta_t$. \cite{bl} It is this
$\theta_t$ that determines the width of the peak in $j_c(\theta)$
caused by the defects. When $H_{ax}$ increases, the characteristic
tilt of the flux lines in the strip, $H_{ax}/H_z$, also increases,
and when the tilt exceeds $\theta_t$, pinning by the columnar
defects becomes ineffective.

\section{How to reconstruct $j_c$ from $H_z(x)$}

We now discuss the problem of the reconstruction of $j_c({\bf H})$
from experimental magnetic-field profiles. Various procedures were
elaborated \cite{R,JH,Gr,W,J,G,Joo,Kr} which  enable one to
extract the spatial distribution of the sheet current $J$  from an
experimental $H_z$ profile measured at the upper surface of a thin
flat superconductor. In particular, for the case of the strip,
equation (\ref{1}) can be explicitly inverted:\cite{JH}
\begin{equation} \label{13}
 J(x)={2\over \pi }
 \int_{-w}^{w}\!\!{H_z(v)-H_{az}\over v-x}
 \left ( {w^2-v^2\over w^2-x^2}\right )^{1/2}dv .
\end{equation}
Eliminating $x$ from $H_z(x)$ and $J(x)$ (in the region of a
nonzero $H_z$), one can find the function $J_c(H_z)$ from
experimental data; see, e.g., Ref.~\onlinecite{ita}.

If the extracted $J_c(H_z)\approx \ $const., this means that the
Bean model is valid, and $j_c=J_c/d$. It is this formula that is
commonly used to find $j_c$ in experiments. However, this simple
approach in general is {\it not correct} when $J_c$ depends on
$H_z$. It is necessary to distinguish between two situations. For
the case of weak collective pinning by point defects in the small
bundle pinning regime, $j_c$ depends only on the combination
$|H|\cos\theta=H_z$. \cite{bl} Then, the critical current density
is constant across the thickness of the strip, and one still has
\begin{equation}\label{14}
  j_c(H_z)={J_c(H_z)\over d},
\end{equation}
i.e., the usual formula is valid. On the other hand, when $j_c$
depends on $\theta$ and $|H|$ separately, and when to the first
approximation one can assume that $j_c=j_c(\theta)$ (see Sec.\
III), formulas (\ref{7}) should be used to reconstruct the
$j_c(\theta)$ from $J_c(H_z)$ obtained at $H_{ax}=0$. In order to
distinguish between these two situations, i.e., to choose between
formulas (\ref{7}) and (\ref{14}), one can use the magnetic field
profiles obtained in oblique magnetic fields.

When $j_c=j_c(H_z)$, the magnetic-field profiles in oblique
magnetic fields depend only on $H_{az}$ and are the same for any
scenario of switching on ${\bf H}_a$. On the other hand, the
results of Sec.~III show that another situation occurs when $j_c$
is not only a function of $H_z$. In this case the penetration
depth of $H_z$ depends on $H_{ax}$. Moreover, different scenarios
can lead to different profiles $H_z(x)$.

When the magnetic-field profiles in oblique magnetic fields show
that the dependence $j_c=j_c(H_z)$ cannot explain the experimental
data, one can apply formulas (\ref{7}) to extract $j_c(\theta)$
from the $H_z$ profiles at $H_{ax}=0$. If the sample is not too
thick, $j_cd < H_0$, where $H_0$ is the characteristic scale of
the $|H|$-dependence of $j_c$, the extracted $j_c(\theta)$
approximately gives $j_c(\theta,H=0)$. For thicker samples the
obtained $j_c(\theta)$ can be considered as a first step in
determining $j_c(\theta,|H|)$. Calculating then the magnetic field
profiles in oblique fields and comparing them with appropriate
experimental data, one can verify the assumption $j_c=j_c(\theta)$
(i.e., the fulfilment of $j_cd < H_0$) for the superconductor
under study. If this assumption does not lead to a good agreement
with the data in oblique magnetic fields, one should
supplement the obtained $j_c(\theta)$ with some $|H|$ dependence.
We emphasize that since the function $J_c(H_z,H_{ax})$ extracted
from the magnetic-field profiles in oblique magnetic fields,
depends on two variables $H_z$ and $H_{ax}$, this function, in
principle, enables one to determine $j_c(\theta,|H|)$ depending on
two variables, too. However, in this determination one should
first assume some model of the ${\bf H}$-dependence of $j_c$ and
then calculate $J_c(H_z,H_{ax})$ from Eq.~(\ref{5}); see, e.g.,
Appendix A. On the other hand, when $j_cd < H_0$, formulas
(\ref{7}) give the model-independent $j_c(\theta,0)$ directly from
the experimental data.

 \acknowledgments

  This work was supported by the German Israeli Research
Grant Agreement (GIF) No G-705-50.14/01.

\appendix

\section{$j_c$ depends on $\theta$ and $H_z$}

Here we present some formulas for reconstruction of $j_c({\bf H})$
when the simplest form $j_c=j_c(\theta)$ cannot sufficiently well
describe the experimental magnetic-field profiles in
perpendicular and oblique magnetic fields. We assume that
$j_c({\bf H})$ can be presented in the form: $j_c({\bf
H})=j_c(\theta)/f(H_z)$ where the function $f(H_z)$ takes into
account a possible dependence of $j_c$ on $|H|$. \cite{B1}

Let the functions $J_c(H_z,0)\equiv J_c(H_z)$ and
$J_c(H_z,H_{ax})$ be extracted from the magnetic-field profiles
obtained for the perpendicular and oblique magnetic fields, see
Eq.~(\ref{13}). In other words, these functions are assumed to be
known here. Then, repeating the analysis of Eq.~(\ref{5}) in
Sec.~III, we arrive at the equations:
     \begin{eqnarray} \label{B1}
 j_c(\theta) d = \left (\! J_c(H_z) -H_z{d J_c(H_z) \over d H_z}
 \right ){H_zf(H_z)\over F(H_z)} \,, \nonumber \\
  \tan \theta   = { J_c(H_z) \over 2 H_z } \,, \\
 \ln [f(H_z)]= \int_0^{H_z}\!\! du\left ({F(u)\over u^2}-
  {1 \over u}\right ) \nonumber
     \end{eqnarray}
that generalize the parametric representation of $j_c(\theta)$,
Eqs.~(\ref{7}). Here the function $F(H_z)$ is determined by the
formula:
     \begin{eqnarray}  \label{B2}
 {1\over F(H_z)}&=&{1\over 2t_+}+\sigma {1\over 2t_-},
     \end{eqnarray}
where $\sigma$ is the sign of $[0.5J_c(H_z,H_{ax})-H_{ax}]$, and
$t_+$, $t_-$ are found from the two uncoupled equations,
     \begin{eqnarray}  \label{B3}
 {0.5J_c(H_z,H_{ax})+H_{ax}\over H_z}&=&{J_c(t_+,0)\over 2t_+},
 \nonumber \\
 {|0.5J_c(H_z,H_{ax})-H_{ax}|\over H_z}&=&{J_c(t_-,0)\over 2t_-}.
      \end{eqnarray}
When $F(H_z)=H_z$, one has $f(H_z)=1$, and Eqs.~(\ref{B1}) reduce
to Eq.~(\ref{7}). Note that the function $F(H_z)$ obtained from
Eqs.~(\ref{B2}), (\ref{B3}) has to be practically independent of
$H_{ax}$ for our assumption $j_c({\bf H})=j_c(\theta)/f(H_z)$ to
be justified.

{}


\begin{thebibliography}{}
\bibitem{c1} In principle, the differences between these critical
states can be detected via the $x$ component of the magnetic
moment. \cite{mbi}
\bibitem{c2} Equations (\ref{8}) may be solved, e.g.,
by writing them in the form $f_1=0$, $f_2=0$, $f_3=0$ where $f_i$
are functions of the three independent variables $t_+$, $t_-$,
$J_c(H_z,H_{ax})$ and then minimizing  $f_1^2+f_2^2+f_3^2$
numerically, starting with appropriate initial values. Using the
solution at some $H_z$, $H_{ax}$ as the initial values for nearby
$H_z$, $H_{ax}$, one can find $t_+$, $t_-$, $J_c$ for any $H_z$,
$H_{ax}$, starting with the known solution at $H_{ax}=0$.
\bibitem{c4} The formulas of Ref.~\onlinecite{mbi} can describe the
critical state of the strip in the core region $|x|\le a$ even for
the case $j_c=j_c(\theta)$ since the flux lines are practically
parallel to the strip plane there and $\theta \approx \pi/2$.
Hence, it is sufficient to put $j_c=j_c(\pi/2)$ in the formulas of
Ref.~\onlinecite{mbi} and to use $J(x)$ obtained numerically from
the solution of the critical state problem for the infinitely thin
strip with the appropriate $J_c(H_z)$.
\bibitem{B1} It is clear that a function of $\theta$ and $|H|$
can be represented as a function of $\theta$ and
$H_z=|H|\cos\theta$. However, the essential assumption is that
$j_c(\theta,H_z)$ is a {\it product} of factors depending on
$\theta$ and $H_z$.
\end{thebibliography}
\end{document}